# *Data hiding in complex-amplitude modulation using a digital micromirror device*

SHUMING JIAO,[1,2] DONGFANG ZHANG,[1,2] CHONGLEI ZHANG,[1,*] YANG GAO,[1] TING LEI,[1] XIAOCONG YUAN[1,*]

[1]Nanophotonics Research Center, Shenzhen University, Shenzhen 518060, Guangdong, China
[2]These authors contributed equally to this work
*Corresponding author: clzhang@szu.edu.cn , xcyuan@szu.edu.cn

**Abstract:**
**A digital micromirror device (DMD) is an amplitude-type spatial light modulator. However, a complex-amplitude light modulation with a DMD can be achieved using the superpixel scheme. In the superpixel scheme, we notice that multiple different DMD local block patterns may correspond to the same complex superpixel value. Based on this inherent encoding redundancy, a large amount of external data can be embedded into the DMD pattern without extra cost. Meanwhile, the original complex light field information carried by the DMD pattern is fully preserved. This proposed scheme is favorable for applications such as secure information transmission and copyright protection.**

The complex modulation of a light field [1,2] receives much attention in many optical applications such as holographic 3D display [3,4], optical communications [5], microscopy [6] and scattering imaging [7]. However, spatial light modulators usually can only modulate either the amplitude component or the phase component of a light field, instead of full complex-amplitude modulation. The digital micromirror device (DMD) is a commonly used amplitude-type spatial light modulator and each pixel can represent a binary amplitude "0" and "1". It has advantages of small pixel size, high frame rate, broad bandwidth and low cost. However, an amplitude-type modulation device cannot be directly employed to modulate the phase information. In previous works, a double lens optical setup and corresponding algorithms were proposed to achieve superpixel-based complex-amplitude light modulation with a DMD [1].

On the other hand, information security is a critical issue in the modern society and various data hiding, watermarking and steganography techniques can prevent illegal access of data from unauthorized users [8,9]. The hidden data can be applied for the copyright and ownership tracking of a photograph, a video, a digital hologram or other contents [8]. Some optical systems and principles are employed to secure data security in a physical way [9], in addition to commonly used digital techniques [8]. For example, a hologram has some unique image characteristics compared with a conventional photograph, which can be more favorable for data hiding as a host carrier [10-15]. However, in most digital or optical data hiding methods, the host carrier will be inevitably damaged when a large amount of hidden data is embedded. In this work, the inherent encoding redundancy in superpixel-based DMD complex modulation is employed and the DMD pattern is designed as a high capacity and non-destructive host carrier for external data hiding, shown in Fig. 1.

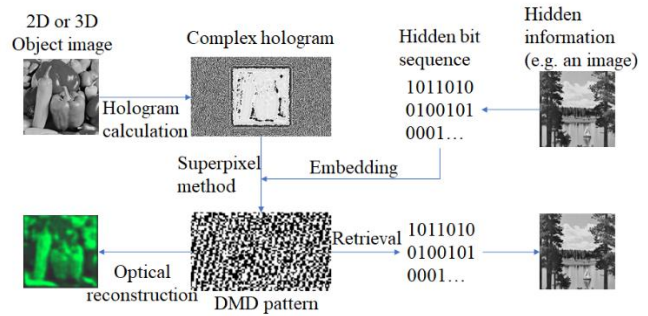

Fig. 1. Proposed data hiding scheme in DMD complex-amplitude modulation.

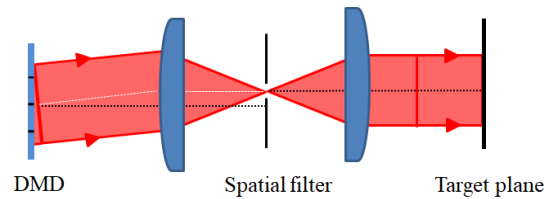

Fig. 2. Optical setup of a superpixel-based complex modulation system with a DMD.

The working principles of a superpixel DMD complex modulation scheme [1] are briefly described as follows. As shown in Fig. 2, the incident light field, modulated by a DMD, passes through a 4-f double-lens configuration. In the Fourier transformed domain, the light field is filtered by a band-pass spatial filter. The spatial filtering operation is equivalent to the multiplication of a periodic phase mask with the DMD plane. Originally each DMD pixel has the same phase value and it will be modulated with a varied phase value in the target plane of this setup.

For example, the phase distributions for each local block containing 4×4 DMD binary pixels (one superpixel) in the target plane are shown in Fig. 3(a). The 16 pixels within a superpixel have 16 different phase values. The 16 binary DMD pixels can be arbitrarily encoded by turning them to be "on" or "off" states. Each super-pixel can represent a complex-amplitude value as the inner product between its 4×4 phase mask (Fig. 3(a)) and its 4×4 binary DMD block pattern (Fig. 4(a)). In Fig. 3, the superpixel can represent a complex-amplitude $exp(j\frac{2\pi}{8}) + \exp(j\frac{4\pi}{8}) + exp(j\frac{16\pi}{8}) = 2.4142\, exp(j\frac{\pi}{4})$ since only these three binary pixels are turned on. For each superpixel, the binary block patterns can be encoded with a total of $2^{16}$ = 65536 possible combinations. However, multiple different encoded block patterns may yield the same superpixel value and there are maximumly 6561 different complex values that each super-pixel can possibly represent. An arbitrary complex value after normalization can usually be approximated to a very close value among the 6561 ones. Consequently, such as a superpixel-based DMD system can perform complex-amplitude light field modulation precisely.

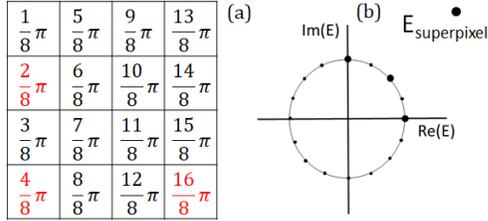

Fig. 3. Superpixel complex-modulation method with a DMD: (a) Periodic random phase mask for one superpixel; (b) Combination of phase values to constitute a complex light field by binary coding.

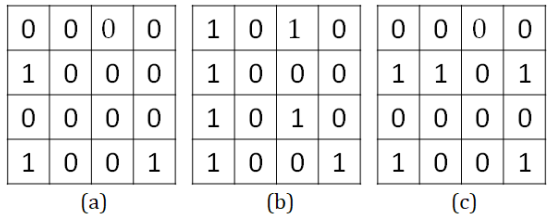

Fig. 4. Three equivalent 4×4 binary coding patterns for one superpixel value $2.4142\, exp(j\frac{\pi}{4})$.

It is a redundancy that multiple different binary block patterns within a superpixel correspond to the same complex-amplitude value. For example in Fig. 4, the three different encoded patterns all correspond to the same complex value $2.4142\, exp(j\frac{\pi}{4})$, like in Fig. 4(c), $exp(j\frac{2\pi}{8}) + exp(j\frac{4\pi}{8}) + exp(j\frac{6\pi}{8}) + exp(j\frac{16\pi}{8}) + exp(j\frac{14\pi}{8}) = 2.4142\, exp(j\frac{\pi}{4})$. In fact, up to 32 different 4×4 DMD block patterns can yield the same super-pixel value $2.4142\, exp(j\frac{\pi}{4})$, from the results of an exhaustive search.

Each superpixel value in a given complex light field can be mapped to the nearest one among the 6561 values stored in the lookup table and a corresponding 4×4 binary block is encoded on the DMD correspondingly. Finally, the entire DMD pattern (e.g. a superpixel array) is generated and can be loaded to a DMD for optical modulation. For one superpixel, if multiple binary patterns are all optimal in producing the target complex value, the best one can be selected randomly ("random strategy") or with some additional rules such as minimizing the total number of "1" ("minimum strategy") or maximizing the total number of "1" ("maximum strategy") in the local block.

In our proposed data hiding scheme, the selection of 4×4 binary block pattern for each superpixel is determined by the encoded hidden information, instead of using "random strategy" or other strategies. The embedded data is in the form of a binary bit sequence and it is assumed the sequence length is within the maximum hiding capacity of the host DMD pattern. First, like the conventional superpixel method [1], a lookup table is prepared in advance. It shall be noted that the lookup table in this work contains more information than the one in the work [1]. The conventional lookup table in [1] only includes one single selected binary block pattern for each of the 6561 complex values. The lookup table in this work includes all the possible 4×4 binary matrices corresponding to each of the 6561 values. It is fixed and independent from the complex light field to be encoded and the hidden data to be embedded. For different hidden bit sequences, different binary block patterns are selected from all the possible ones stored in the lookup table correspondingly to constitute each superpixel in the generated DMD pattern. In the lookup table, each binary block pattern will be assigned with a different index if they correspond to the same complex value. For example, if the complex-amplitude is $2.4142\, exp(j\frac{\pi}{4})$, 32 different binary local block patterns will be labeled as 00000, 00001, …, and 11111.

Then in the encoding of the DMD pattern, each superpixel is processed sequentially. The complex amplitude C of a given superpixel is first mapped to one of the 6561 values, $S_m$ ($1 \leq m \leq 6561$). It is assumed that there are totally $N_m$ different binary local blocks corresponding to $S_m$ and $d = \text{Floor}[log_2(N_m)]$ bits can be hidden in the current superpixel, where Floor[ ] refers to the nearest integer less or equal to the actual value within the bracket. $\text{Floor}[log_2(N_m)]$ bits will be taken from the bit stream to be hidden and the binary local block pattern will be selected based on the index represented by these bits from the lookup table. It shall be noted that the hidden capacity of each superpixel varies depending on $N_m$ and it is possible that no data is hidden in some superpixels. Finally, the entire hidden data sequence can be embedded into the host DMD pattern (or DMD superpixel array). The inverse retrieval of hidden data from the watermarked DMD pattern can be realized in an opposite way. According to the bit values of the 4×4 local binary block for each superpixel, the corresponding index (e.g. a bit string 0101) can be found by searching the lookup table and the index string will be added to the retrieved hidden bit sequence. The flowcharts of hidden data embedding and retrieval are shown in Fig. 5 and Fig. 8.

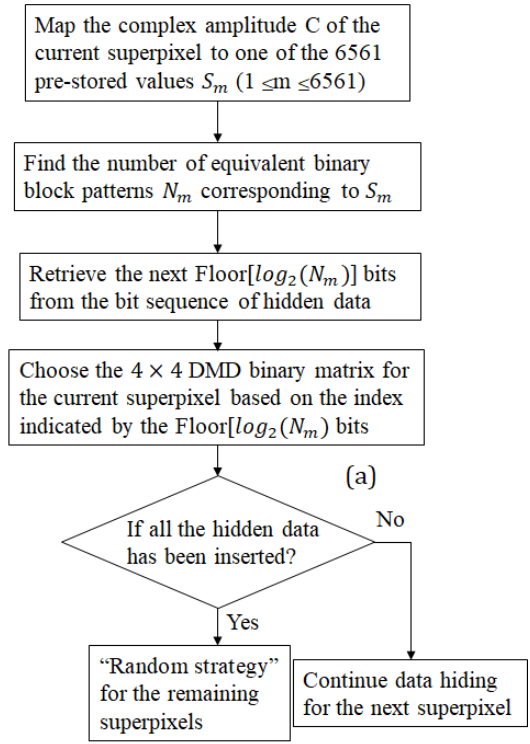

Fig. 5. Flowcharts of our proposed scheme: hidden data embedding.

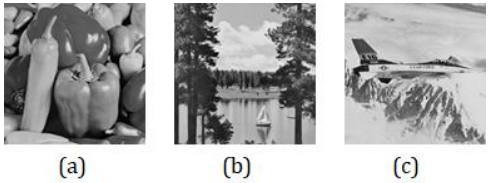

Fig. 6 (a) Object image for the generation of the complex Fresnel hologram; (b) Hidden image $P_1$ and (c) Hidden image $P_2$.

In our experiment, a complex-amplitude hologram H is first calculated from an object image (750 ×750 pixels) shown in Fig. 6(a) by Fresnel transform [16,17]. Then the complex hologram will be converted to a DMD pattern (1080 ×1920 pixels) by the superpixel scheme. There are totally 270 ×480 superpixels and each superpixel contains 4 ×4 DMD pixels. The pixel size is 7.56 μm and the wavelength is 520 nm. The distance between the object plane and the hologram plane is 0.2m. The DMD complex modulation system described above is employed to optically reconstruct the generated hologram H. A VIsionFly 6500 DMD is used in our system. First, based on the conventional superpixel method [1] without data hiding, the generated DMD patterns and the corresponding reconstructed images under "random strategy", "minimum strategy" and "maximum strategy" are compared in Fig. 7(a)-(c) and Fig. 9(a)-(c). The SSIM values of reconstructed images in the simulation are 0.5949, 0.6026 and 0.5999. The three DMD patterns appear to be quite different. But the visual qualities of reconstructed images are very similar and it indicates that different selection strategies of binary block patterns for the same complex amplitude in the superpixel method are almost experimentally identical.

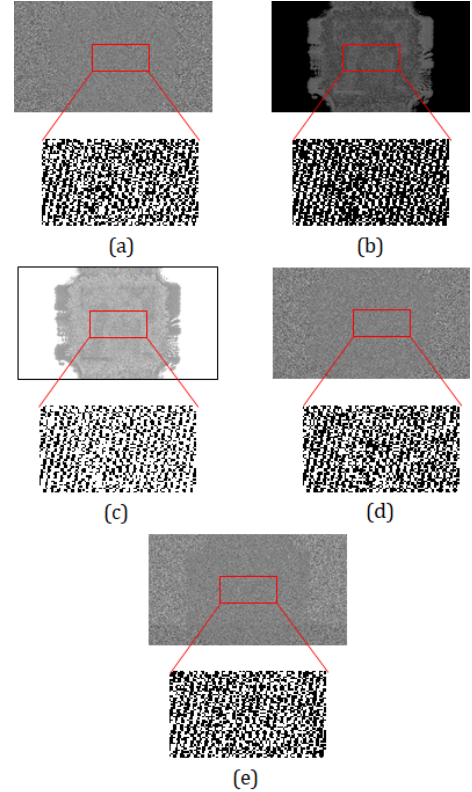

Fig. 7. Generated DMD patterns corresponding to the complex-amplitude hologram without data embedding: (a) random strategy; (b) minimum strategy; (c) maximum strategy; Generated DMD patterns corresponding to the complex-amplitude hologram with: (d) Image $P_1$ hidden; (e) Image $P_2$ hidden.

Our proposed data hiding scheme is experimentally verified as follows. Two grayscale images (256 ×256 pixels) $P_1$ and $P_2$ are used as the hidden images, shown in Fig. 6(b) and 6(c). Each grayscale pixel value in the hidden images is represented by 8 bits and each hidden image is transformed to a binary sequence of length 256 ×256×8 = 524288 bits. To enhance security, each bit sequence is randomly permutated with an encryption key. The maximum hiding capacity for H is 656767 bits in our scheme, which is the summation of the number of bits that can be hidden in each superpixel. Two different DMD patterns $D_1$ and $D_2$ are generated for the same complex hologram H but with $P_1$ embedded and $P_2$ embedded respectively. After embedding, the hidden images can be retrieved from the watermarked DMD patterns in an inverse way. The hidden bit sequence can be first retrieved and the grayscale pixel values of the hidden image can be recovered from every 8 bits in the sequence.

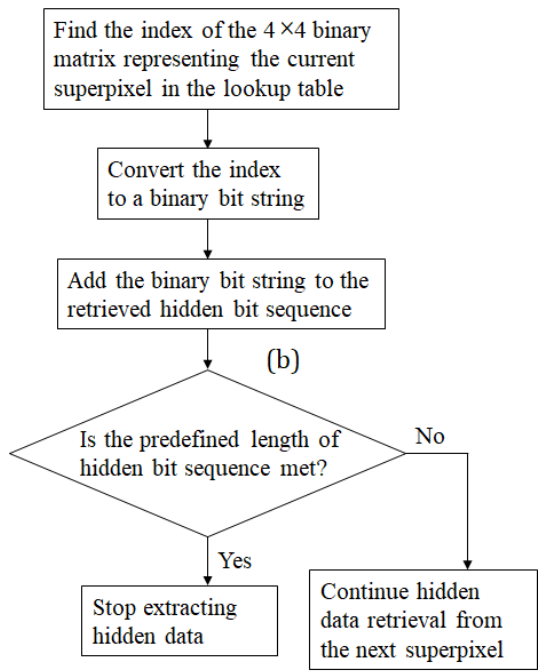

Fig. 8. Flowcharts of our proposed scheme: hidden data retrieval.

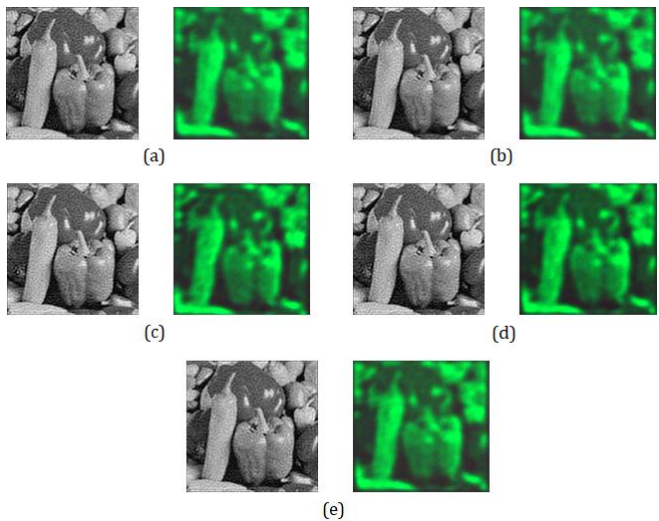

Fig. 9. Reconstructed images (left: simulation; right: experiment) from the DMD patterns without data hiding: (a)random strategy; (b)minimum strategy; (c)maximum strategy; Reconstructed images (left: simulation; right: experiment) from the DMD patterns with (d) Image $P_1$ hidden; (f) Image $P_2$ hidden.

From the reconstruction results of the watermarked DMD patterns (or DMD superpixel array), it can be observed that the image qualities in Fig. 9(d)-(e) are very similar to the ones in Fig. 9(a)-(c). The SSIM values of two reconstructed images in the simulation are 0.5913 and 0.5950. In addition, the hidden images are not observable from the DMD patterns in Fig. 7(d)-(e). It is hard for attackers to identify whether a DMD pattern contains hidden information (e.g. Fig. 9(d)-(e)) or not (e.g. Fig. 9(a)). It can be concluded that the embedding of hidden images does not degrade the complex modulation capability of the original system.

In summary, the inherent encoding redundancy in the superpixel-based DMD complex modulation system is employed for data hiding. Based on the hidden bit sequence, each DMD binary local block is selected among multiple equivalent ones corresponding to the same superpixel value. The external hidden data, such as a grayscale image, can be embedded into the local superpixel blocks in the DMD pattern and retrieved in an inverse manner. The host DMD pattern has a high hiding capacity and will not be damaged after watermarking, which is favorable for applications such as secure information transmission and copyright protection.